\documentclass[12pt]{article}

\usepackage{graphicx}
\usepackage{amssymb,amsfonts,amsmath}
\usepackage[superscript,biblabel]{cite}
\usepackage{times}
\usepackage{fullpage}
\usepackage{color}
\usepackage{bm}

\usepackage{hyperref}

\renewcommand{\baselinestretch}{1.5}

\newcommand{\Rey}{\ensuremath{R}}

\title{The rise of fully turbulent flow} 

\author{
Dwight Barkley$^{1}$,
Baofang Song$^{2,3}$,
Vasudevan Mukund$^{2}$,
Gr\'egoire Lemoult$^{2}$, \\
Marc Avila$^{4}$, and
Bj\"orn Hof$^{2\ast}$
}

\date{6 June, 2015}

\begin{document}

\maketitle

\noindent
$^{1}$ 
Mathematics Institute, University of Warwick, Coventry, CV4 7AL, United Kingdom

\noindent
$^{2}$ 
Institute of Science and Technology Austria, 3400 Klosterneuburg, Austria

\noindent
$^{3}$ 
Institute of Multiscale Simulation, Friedrich-Alexander-Universit\"at, Erlangen, Germany

\noindent
$^{4}$ 
Institute of Fluid Mechanics, Friedrich-Alexander-Universit\"at, Erlangen, Germany

\noindent
$^\ast$ email: bhof@ist.ac.at

\vspace{2cm}
\begin{center}
This manuscript is the original submission to Nature. The final (published)
version can be accessed at
\href{http://www.nature.com/nature/journal/v526/n7574/full/nature15701.html}
     {http://www.nature.com/nature/journal/v526/n7574/full/nature15701.html}
\end{center}


\newpage


{\bf Over a century of research into the origin of turbulence in wallbounded
  shear flows has resulted in a puzzling picture in which turbulence appears
  in a variety of different states competing with laminar background flow
  \cite{reynolds1883,Coles:1962,wygnanski1973,
    sreenivasan1986,nishi2008,mullin2011}. At slightly higher speeds the
  situation changes distinctly and the entire flow is turbulent.  Neither the
  origin of the different states encountered during transition, nor their
  front dynamics, let alone the transformation to full turbulence could be
  explained to date.  Combining experiments, theory and computer simulations
  here we uncover the bifurcation scenario organising the route to fully
  turbulent pipe flow and explain the front dynamics of the different states
  encountered in the process.  Key to resolving this problem is the
  interpretation of the flow as a bistable system with nonlinear propagation
  (advection) of turbulent fronts.  These findings bridge the gap between our
  understanding of the onset of turbulence \cite{kavila2011} and fully
  turbulent flows \cite{pope2000, schlichting2000}.
}


The sudden appearance of localised turbulent patches in an otherwise quiescent
flow was first observed by Osborne Reynolds for pipe flow \cite{reynolds1883}
and has since been found to be the starting point of turbulence in most shear
flows \cite{emmons1951,Coles:1962,coles1965,
  sreenivasan1986,lundbladh1991,tillmark1992,cros2002,lemoult2013}.
Curiously, in this regime it is impossible to maintain turbulence over
extended regions as it automatically \cite{moxey2010,samanta2011} reduces to
patches of characteristic size, called puffs in pipe flow (see Fig.~1a).
Puffs can decay, or else split and thereby multiply. Once the Reynolds number
$\Rey>2040$ the splitting process outweighs decay, resulting in sustained
disordered motion.\cite{kavila2011} Although sustained, turbulence at these
low $\Rey$ only consists of puffs surrounded by laminar flow (Fig.~1a) and
cannot form larger clusters \cite{hof2010,samanta2011}.

At larger flow rates, the situation is fundamentally different: once
triggered, turbulence aggressively expands and eliminates all laminar motion
(Fig.~1b).  Fully turbulent flow is now the natural state of the system and
only then do wallbounded shear flows have characteristic mean properties such
as the Blasius or Prandtl-von Karman friction laws \cite{schlichting2000}.
This rise of fully turbulent flow has remained unexplained despite the fact
that this transformation occurs in virtually all shear flows and generally
dominates the dynamics at sufficiently large Reynolds numbers.

\renewcommand{\baselinestretch}{1.0}
\begin{figure}
\begin{center}
\includegraphics[width=4.0in]{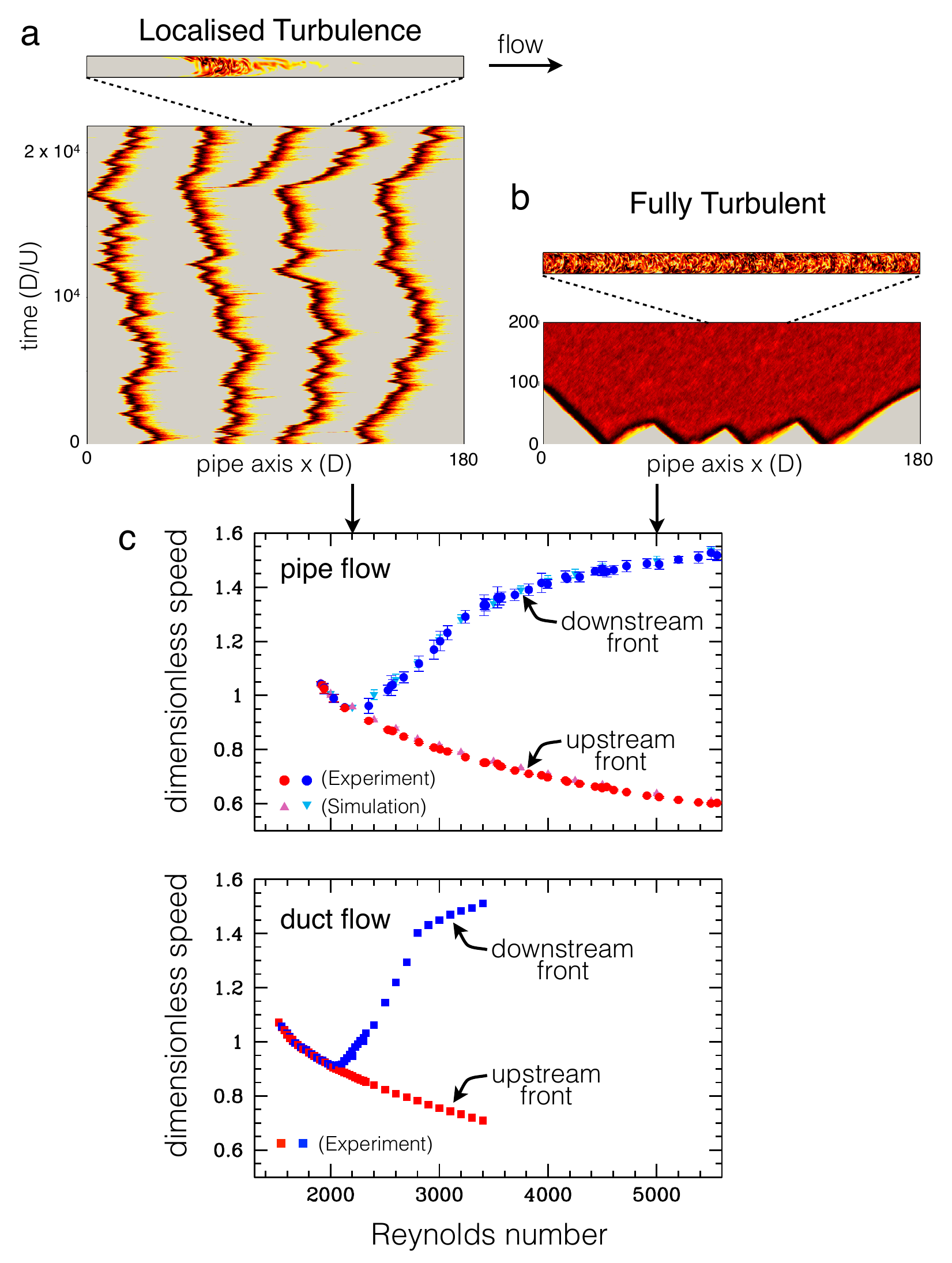} 
\end{center}
\caption{{\bf Localised and fully turbulent flow.}  {\bf a, b}, numerical
  simulations of pipe flow illustrate the distinction between ({\bf a})
  localised turbulence, at $\Rey=2200$ and ({\bf b}) fully turbulent flow, at
  $\Rey=5000$. In each case, the flow is initially seeded with localised
  turbulent patches and the subsequent evolution is visualised via space-time
  plots in a reference frame co-moving with structures.  Colours indicate the
  value of $\sqrt{u_r^2+u_{\theta}^2}$. Further highlighting the distinction
  between cases, shown at the top are cross sections of instantaneous flow
  within the pipe. A 35D section is shown with the vertical direction
  stretched by a factor of 2.  While the protocol used here is seeding the
  flow with localised patches of turbulence, the fundamental distinction
  between localised and fully turbulent flow is independent of how turbulence
  is triggered \cite{moxey2010}.  {\bf c}, Speeds of turbulent-laminar fronts
  as a function of Reynolds number for pipe flow and duct flow. A speed
  difference between the upstream and downstream fronts corresponds to
  expanding turbulence, and asymptotically, to a fully turbulent flow.  }
\end{figure}
\renewcommand{\baselinestretch}{1.5}

A classical diagnostic for the formation of turbulence
\cite{Lindgren1957,Coles:1962,wygnanski1973,
  sreenivasan1986,nishi2008,Duguet:2010} is the propagation speed of the
upstream and downstream fronts for a turbulent patch.  We have carried out
such measurements for pipe and square duct flow (Fig.~1c), focusing on the
regime where turbulence first begins to expand. In both experiments, fluid
enters the conduit through a smoothly contracting inlet, ensuring that without
external perturbations flows are laminar over the entire Reynolds number
regime investigated. Turbulence is triggered $120 D$ from the inlet by a
short-duration, localised perturbation. A pressure sensor at the outlet
determines the subsequent arrival of first the downstream and then the
upstream turbulent-laminar front. Speeds are averaged over many realisations
for each $\Rey$, giving a total travel distance of typically $5\times10^4 D$.
As an independent verification, speeds in pipe flow were determined through
extensive direct numerical simulations in pipes of length $180D$, with
averaging over large ensembles.

In both flows, initially the speeds of the downstream fronts are
indistinguishable from the upstream ones, signalling localised turbulence. For
$\Rey \gtrsim 2250$ in pipe flow and $\Rey \gtrsim 2030$ in duct flow, the
downstream speed increases with $\Rey$, marking the point where turbulence
begins to aggressively invade the surrounding fluid.  With further increases
in $\Rey$ the downstream front speeds exhibits complex changes of curvature as
a function of $\Rey$.  Surprisingly, the spreading of turbulence shows neither
a square-root scaling nor an exponent associated with a percolation type
processes, as proposed in earlier studies~\cite{sreenivasan1986,sipos2011}.
The speed of the downstream spreading has a far more complex behaviour than
these theories imply.


In a theoretical approach \cite{barkley2011,barkley2011b,Barkley:2012}, puffs
in pipe flow were categorised as localised excitations analogous to action
potentials in axons and from this numerous features of puff turbulence were
captured. However, in that model the transition leading to an expanding state
is first-order (discontinuous), not reflecting the observed continuous
behaviour at the onset of fully turbulent flow (Fig.~1c).  This model did not,
however, include nonlinear advection, a feature intrinsic to fluid dynamics.
We have devised an extended model incorporating advective nonlinearity that
fully captures the sequence encountered in the route to fully turbulent flow.
The model is
\begin{equation}
q_t + (u - \zeta) q_x = f(q,u) +  D q_{xx}, \quad  
u_t + u u_x = \epsilon g(q,u)
\end{equation}
where,
\begin{equation}
f(q,u) = q \left( r + u - 2 - (r + 0.1) (q - 1)^2  \right), \quad
g(q,u) = 2 - u + 2 q (1 - u)
\end{equation}
Subscripts denote derivatives.  Variables $q$ and $u$ depend only on the
streamwise coordinate $x$ and time $t$.  $q$ represents the turbulence level
while $u$ represents the centreline velocity and plays two important roles. It
accounts for nonlinear advection in the streamwise direction and it models the
state of the shear profile: $u=2$ for parabolic flow and $u<2$ for plug flow.
The functions $f(q,u)$ and $g(q,u)$ capture the known interplay between
turbulence (the excited state), and the shear profile\cite{barkley2011,
  wygnanski1973}.  Parameter $r$ plays the role of Reynolds number, $\zeta$
accounts for the fact that turbulence is advected more slowly than the
centreline velocity, $D$ controls the coupling strength of the turbulent
field, and $\epsilon$ sets the timescale ratio between fast excitation of $q$
and slow recovery of $u$ following relaminarisation.  The fast scale of $q$
and cubic nonlinearity in $f(q,u)$ are motivated by known upper- and
lower-branch exact coherent structures in shear flows \cite{eckhardt2007,
  gibson2008, kawahara2011}.  (See SI for details.)

\renewcommand{\baselinestretch}{1.0}
\begin{figure}
\begin{center}
\includegraphics[width=4.5in]{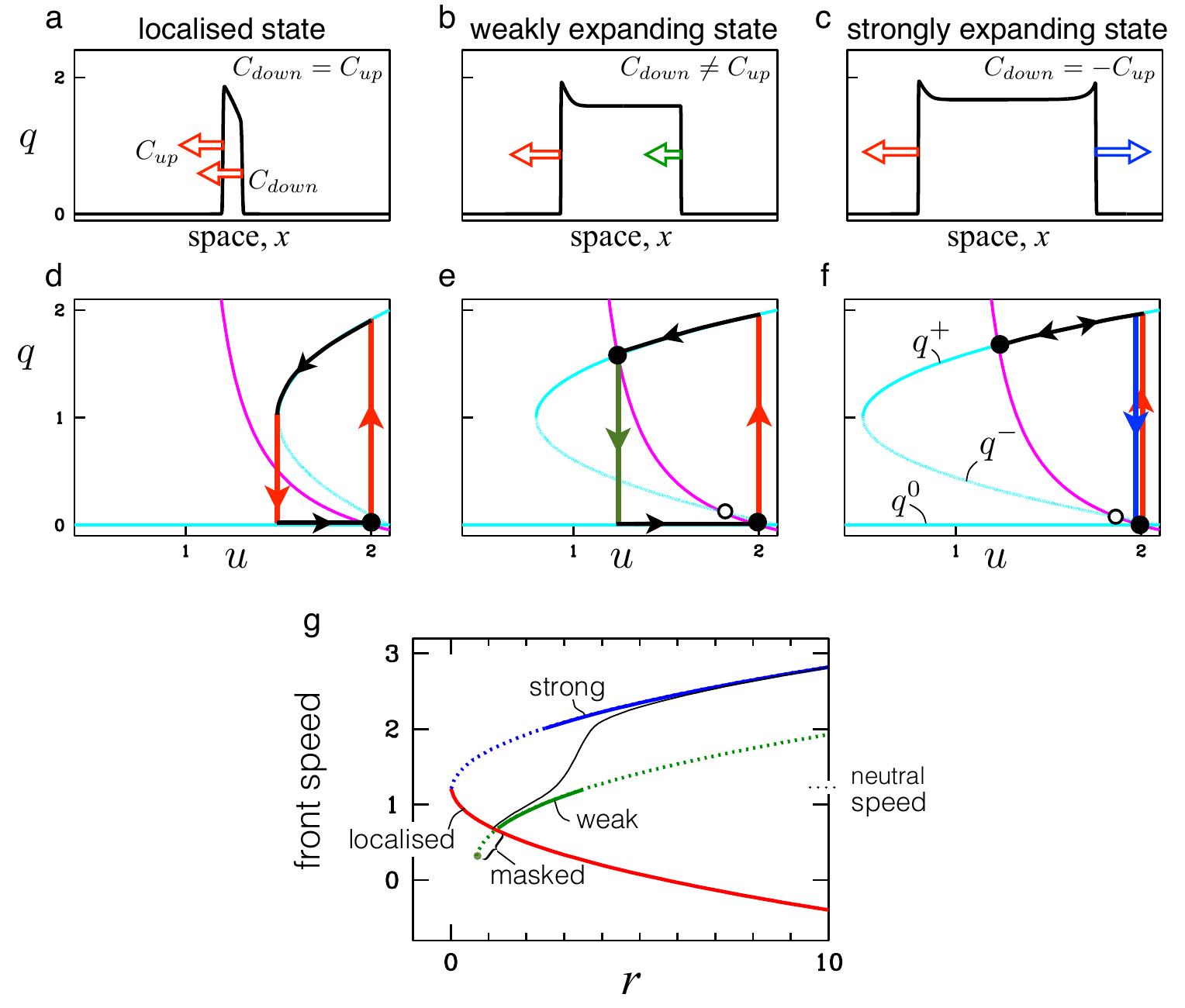} 
\end{center}
\caption{{\bf Model predictions in the asymptotic limit of sharp fronts.}
  {\bf a}-{\bf c}, Three distinct types of predicted states. $C_{up}$ and
  $C_{down}$ are the upstream and downstream front speeds. {\bf d}-{\bf f},
  corresponding states seen in the local phase plane with arrows indicating
  increasing space coordinate $x$, (not time).  The $q$-nullcline, (cyan,
  given by $f(q,u)=0$, for clarity only labelled in {\bf f}), has
  three branches since $f$ is cubic in $q$: stable laminar $q^0=0$ and upper
  $q^+$ branches and an unstable branch $q^-$ separating the two. The
  $u$-nullcline (magenta, given by $g(q,u)=0$), describes the decrease in the
  centreline velocity in the presence of turbulence and its recovery in the
  absence of turbulence.  Fronts are jumps between stable branches of the $q$
  nullcline.  In all cases, the upstream front is a transition from laminar
  flow (the equilibrium at $u=2,q=0$) to the upper branch. The cases are
  distinguished by the downstream front.  In {\bf a}, {\bf d} the system is
  excitable and the downstream transition, from $q^+$ to $q^0$, is
  unrestricted by the upper branch and the speed will be selected to match the
  upstream front $C_{down}=C_{up}$, thus giving localised turbulence.  In {\bf
    b}, {\bf e} the system has become bistable with the formation of an upper
  branch steady state. Evolution on the upper branch is restricted by this
  state so the downstream front speed may no longer be able match the upstream
  front speed: $C_{down} \ne C_{up}$ in general. The turbulent patch
  expands. In {\bf c}, {\bf f} the upstream and downstream fronts have the
  same character but move in opposite directions $C_{down} = -C_{up}$ in a
  reference frame moving at the neutral speed. We refer to the downstream
  fronts in {\bf b}, {\bf e} as weak fronts and those in {\bf c}, {\bf f} as
  strong fronts.  {\bf g}, Front speeds as a function of model Reynolds number
  $r$. The nominal critical point for the onset of fully turbulent flow is
  masked.  The neutral speed is the speed about which the upstream and
    strong downstream fronts speeds are symmetric. At finite $\epsilon$ the
    transition from weak to strong scaling is continuous (black curve).
}
\end{figure}
\renewcommand{\baselinestretch}{1.5}

To elucidate the core of the transition from localised to expanding
excitations, and to identify the different states occurring in the process, we
carry out a standard asymptotic analysis~\cite{Rinzel:1982,Tyson:1988} in the
limit of sharp fronts ($\epsilon \to 0$).  Three distinct turbulent structures
are predicted: a localised state (Fig.~2a), an asymmetric expanding state
(Fig.~2b), and a symmetric expanding one (Fig.~2c).  The essence of each state
is clearly seen in the local phase plane (Fig.~2d,e,f).  Equilibrium points
are located at the intersections of the $q$ and $u$ nullclines (curves where
derivatives of $u$ and $q$ are zero).  For low values of $r$ (Fig.~2d) the
only equilibrium is $(u=2,q=0)$, corresponding to parabolic laminar flow.
Nevertheless, the system can be excited locally; when perturbed the state
jumps to the upper branch $q^+$. This forms the upstream laminar-to-turbulent
front. On the upper branch, $\dot u < 0$ and $u$ decreases to a point where
turbulence is not maintained and the system jumps back to $q=0$, forming the
downstream front. The downstream front simply follows the upstream one by a
fixed distance, thus creating a localised excitation: a puff in pipe flow
analogous to an action potential in excitable media
\cite{Rinzel:1982,Tyson:1988,Barkley:2012}.

For larger values of $r$, a second stable equilibrium appears (uppermost
intersection of the nullclines in Fig.~2e,f), and the system is now
bistable. Here fully turbulent flow begins to arise. The downstream front lags
the upstream front giving rise to a growing turbulent region
between. Initially the downstream front is tame and expansion is modest. The
drop from $q^+$ to $q=0$ occurs directly from the upper equilibrium
(Fig.~2d,e).  We refer to this as the {\em weak front state}. For larger $r$
the weak front becomes unstable giving rise to the final state, a much more
rapidly expanding {\em strong front state} (Fig.~2c,f).  The strong downstream
front is the mirror image of the upstream front with an overshoot of $q$ in
space and a drop from $q^+$ to $q=0$ at $u=2$ in the phase plane.  The
downstream speed is opposite the upstream speed with respect to what we term
the {\em neutral speed}.

Before comparing the model to the experimental data, we discuss features of
the front-speed scaling that are intrinsic to this model.  Figure~2g shows
front speeds of the three states. Starting at low $r$, excitations are
strictly localised and their speed monotonically decreases with $r$ (red
curve). Expanding turbulence is first encountered when this curve is
intersected by the weak-front curve (green). Interestingly, the turbulent
state (upper fixed point) bifurcates at lower $r$, but initially the
downstream speed is smaller than the upstream one resulting in a contraction
back to a localised excitation.  Thus onset of bistability and the expansion
do not coincide, masking the transition and resulting in a non-standard front
speed scaling. (In contrast to the case without nonlinear advection shown in
Fig.~S1a).  The strong front (blue) is stable at slightly higher $r$ (solid
portion of curve) and is perfectly symmetric to the downstream front (red)
about the neutral speed.  In the asymptotic limit ($\epsilon \to 0$), weak and
strong fronts co-exist over a range of $r$, but for finite $\epsilon$ the
front speed continuously varies from a weak to increasingly strong front
(solid black curve). During this adjustment the front speed exhibits two
curvature changes. This, together with the eventual approach to the upper
branch of the parabola are distinct signatures of the scenario described by
this model.


Using the theoretical model as a guide, we collapse the measured fronts speeds
from pipe and duct flow and compare them directly with theory (Fig.~3a).
Initially, at lower values of $\Rey$, turbulent excitations are localised (as
illustrated for duct flow in Fig.~3b and pipe flow in Fig.~3e) and the front
speed data from both flows collapse and agree very well with the parabolic
scaling predicted by the model asymptotics (solid red curve).  At $\Rey
\approx 2250$ in pipe flow and $\Rey \approx 2030$ in duct flow, expansion
begins with the formation of the weak downstream front (illustrated for duct
and pipe flow in Figs.~3c and 3f respectively). While upstream fronts of both
data sets continue following the simple asymptotic form, the weak downstream
fronts do not collapse to a single scaling. Nevertheless, with appropriate
choices of parameters $\zeta$ and $\epsilon$, the model precisely captures the
two curvature changes (solid black curves) encountered as each flow
continuously adjusts from the weak front (green dashed) to ultimately the
strong front (blue dashed), corresponding to the emergence of the final strong
front state (Figs.~3d and 3g).  As the downstream front approaches the scaling
given by the strong-front asymptotics, its speed indeed forms a parabola with
the upstream front speed, a feature systematically overlooked in previous
studies.

\renewcommand{\baselinestretch}{1.0}
\begin{figure}
\begin{center}
\vspace{-12mm}
\includegraphics[width=3.4in]{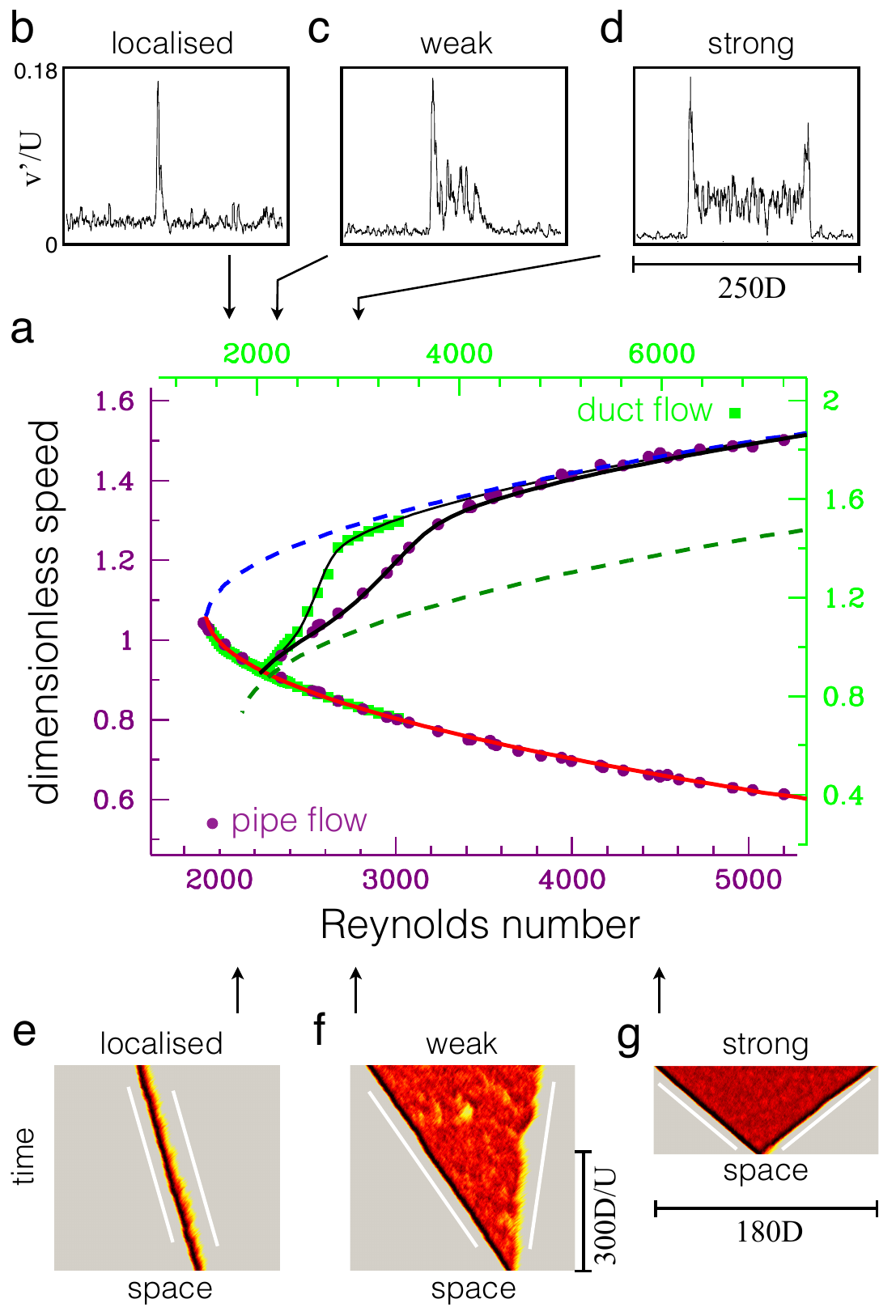} 
\end{center}
\caption{{\bf The rise of fully turbulent flow.}  {\bf a}, front speeds as a
  function of Reynolds number for pipe and duct flow. Points are experimental
  results from Fig.~1c. Red, blue, and dark green curves are front speeds in
  the asymptotic limit of sharp fronts (as in Fig.~2g). The only model
  parameter used for these curves is $D=0.13$.  Black curves are the
  downstream front speed at finite front width ($\epsilon=0.2$, $\zeta=0.79$
  for pipe flow and $\epsilon=0.11$, $\zeta=0.56$ for duct flow).  The
  distinct weak and strong asymptotic branches (dashed) form the skeleton for
  the formation of fully turbulent flow, while at finite front width the model
  captures the complex behaviour of front speeds as a smooth switching between
  the asymptotic branches.  {\bf b}-{\bf d}, crossstream velocity fluctuations
  for the three front states in a square duct: localised puff ($\Rey=1700$),
  expanding turbulence with a weak downstream front ($\Rey=2300$) and the
  strong front state ($\Rey=3000$), exhibiting the characteristic energy
  overshoot at the downstream edge\cite{nishi2008,Duguet:2010}. {\bf e}-{\bf
      g}, Space-time plots from simulations of pipe flow at $\Rey=2000$,
  $\Rey=2800$, and $\Rey=4500$. $\sqrt{u_r^2+u_{\theta}^2}$ is plotted in the
  reference frame moving at the neutral speed.  White lines indicate front
  speeds from the model converted to physical units.  At $\Rey=2000$
  turbulence is localised with equal upstream and downstream front speeds. At
  $\Rey=4500$ turbulence expands with a strong downstream front and the
  long-time flow is fully turbulent.  Note the upstream and downstream fronts
  have the same character (compare with the symmetric overshoot in Fig.~2d)
  and the spreading is symmetric in the neutral reference frame.  At
  $\Rey=2800$ the downstream front moves at a speed between the weak and
  strong branches and exhibits some characteristics of both fronts as it
  fluctuates. This, as well as the intermittent laminar patches appearing
  within the turbulent flow are typical of turbulence as fully turbulent flow
  first arises.
}
\end{figure}
\renewcommand{\baselinestretch}{1.5}

There are two features of pipe and duct turbulence that the model does not
capture.  Both originate from stochastic fluctuations within turbulence and
are most prevalent when turbulence first begins to expand (Fig.~3c, 3f).
Fronts fluctuate, especially the downstream front, and it is common for the
system to exhibit sometimes a strong and sometimes a weak downstream front.
The bifurcation scenario predicted by the model is only recovered in average
quantities.  Likewise, turbulence in this regime is not always uniform, but
commonly contains intermittent laminar pockets \cite{moxey2010,avila2013}.

Nevertheless, the simplicity of the model permits investigation of new
phenomena associated with fully turbulent flow. In the model the creation of
extended turbulent regions hinges on the upper intersection of the $q$ and $u$
nullclines.  The variable $u$ corresponds to profile shape in pipe flow and
this can be manipulated. As suggested by the model, deformation of the profile
can indeed eliminate the upper turbulent fixed point and hence relaminarise
turbulent flow (demonstrated in the SI), offering a very simple and robust way
to control turbulence and to reduce frictional drag.

While in recent years much progress has been made in our understanding of how
turbulence in wallbounded flows is formed from unstable invariant solutions
\cite{eckhardt2007, gibson2008, kawahara2011} at moderate $\Rey$, little to no
progress has been made in connecting this transitional regime to studies of
high-$\Rey$ turbulence. Explaining the origin of the fully turbulent state
is a decisive step towards connecting these regimes and paves the way
for a bottom up approach to turbulence.

\bigskip
\bigskip
\bigskip
\bigskip

\setlength{\parindent}{0in}

\textbf{Acknowledgements} \\
  We thank A.P.~Willis for sharing his hybrid spectral finite-difference code,
  Xiaoyun Tu for helping to set up and test the experiment.
  We acknowledge the Deutsche Forschungsgemeinschaft
  (Project No. FOR 1182), and the European Research Council under the European
  Union's Seventh Framework Programme (FP/2007-2013)/ERC Grant Agreement
  306589 for financial support.  Baofang Song acknowledges financial support
  from the Chinese State Scholarship Fund under grant number 2010629145.  We
  acknowledge computing resources from GWDG (Gesellschaft f\"ur
  wissenschaftliche Datenverarbeitung G\"ottingen) and the J\"ulich
  Supercomputing Centre (grant HGU16) where the simulations were performed.

\textbf{Author contributions} \\ V.M., G.L.\ and B.H.\ designed and performed the
experiments. V.M.\ analysed the experimental results.  B.S.\ and
M.A.\ designed and performed computer simulations of the Navier--Stokes
equations, analysed the results and generated corresponding visualisations.
D.B. performed the theoretical analysis. DB. B.S., V.M., G.L., M.A and B.H. wrote
the paper.

\textbf{Correspondence} \\
  Supplementary information is available in the online
  version of the paper.  Correspondence and requests for materials should be
  addressed to D.B.~(email: D.Barkley@warwick.ac.uk) for theoretical aspect or
  B.H.~(email: bhof@ist.ac.at) for experimental aspects.

\textbf{Competing Interests} \\
  The authors declare that they have no competing financial interests.

\newpage

\setcounter{table}{0}
\renewcommand\thetable{S\arabic{table}}
\setcounter{figure}{0}
\renewcommand\thefigure{S\arabic{figure}}
\renewcommand{\baselinestretch}{1.5}

\section{Materials and methods}

Speeds of laminar-turbulent fronts have been measured in detailed experiments
and highly resolved computer simulations. In both cases long observation times
were necessary to average out stochastic fluctuations that, although intrinsic
to turbulence, may disguise the underlying transition scenario. 

\subsection{Pipe experiments}

Experiments were carried out in a $D=10mm$ ($\pm 0.01mm$) diameter pipe with a
length of $1500 D$.  The $15m$ long pipe was assembled on a straight aluminium
base and made of precision bore glass tubes with lengths between $1m$ and
$1.2m$. Customised connectors made from perspex allowed an accurate fit of the
pipe segments. A specially made pipe inlet consisting of several meshes and
a smooth convergence from a $100mm$ wide section to the $10mm$ pipe was used
to avoid inlet disturbances and eddie formation (see Samanta {\em et al.}
\cite{samanta2011} for details).  In this way the water flow could be held
laminar for $\Rey$ in excess of 8000. 

The laminar flow was then left to develop its parabolic profile over a length
of 200$D$. At this downstream location, the flow could be perturbed by an
impulsive jet of water injected (for 10$ms$) through a 1$mm$ hole in the pipe
wall. The perturbed flow was then left to grow into a turbulent patch over the
next 250 pipe diameters and at this location (i.e. 450$D$ form the inlet), a
pressure sensor recorded the arrival of the upstream and downstream laminar
turbulent interfaces. A second sensor was located a further 1000$D$ downstream
(i.e.\ 50$D$ upstream of the pipe exit) again determining the arrival of the
interfaces so that the average interface speed over the intermediate stretch
of 1000$D$ was measured.  At each Reynolds number, the measurement of the
interface velocity was repeated 10 times.

The flow was gravity driven from a reservoir at a fixed height above the pipe
exit. Since during the course of a measurement the turbulent fraction in the
pipe is increasing, the overall drag in the pipe also increases (turbulent
flow has a higher skin friction than laminar flow). This unavoidably leads to
a drop in the flow rate (and hence $\Rey$) during a measurement. In order to
minimise this effect, a large reservoir height was chosen; in this case 23$m$
above the pipe exit. A precision valve positioned directly in front of the
pipe inlet was used to adjust the flowrate and hence to select $\Rey$. For the
Reynolds number regime investigated here the total pressure drop across the
pipe is much smaller than the 23$m$ water head and most of the pressure drop
occurs across the valve. The increase in drag caused by the expansion of
turbulence is only a small fraction (smaller than $0.5\%$ of the overall
pressure drop) and hence even at the highest Reynolds numbers investigated
flow rates were constant to within better than $0.5\%$ throughout the
measurement.

\subsection{Duct experiments}

Experiments were carried out in a square duct with width $h=5mm$ and a length
of $1200 h$ ($6 m$) . The duct was made of 8 perspex sections precisely
machined to an accuracy of $\pm0.01 mm$. They were assembled and mounted
straight together on an aluminium frame. A well designed entrance section
consisting of a honeycomb and a convergent section, with an area ratio of 25,
allowed the flow to remain laminar up to, at least, $\Rey=5000$, where
$\Rey=u_{\rm{bulk}}h/\nu$.

The flow was gravity driven from a reservoir at a fixed height and water was
used as working fluid. Analogous to the pipe experiment a precision valve was 
positioned directly in front of the duct and was used to set the flowrate. 
The pressure drop across the valve was considerably larger than that across 
the pipe. The temperature of the water was
controlled by means of an heat exchanger that the water had to pass before 
entering the pipe. Overall an accuracy in $\Rey$ of better than $0.5\%$ 
was achieved for the investigated Reynolds number regime.

The flow was perturbed by injecting water through a $0.5 mm$ hole drilled in
one wall of the duct $120h$ downstream from the inlet.  The duration of the
perturbation was varied with $\Rey$ so that in dimensionless units it
corresponded to $5h/U$. The evolution of the perturbation was then monitored at
five locations where the pressure was recorded. The pressure sensors were
positioned at $100h$, $400h$, $600h$, $800h$ and $1000h$ downstream of the
perturbation point. Sensors measured the pressure difference over $10h$ along
the duct. The arrival times of both interfaces were detected at each location
and the overall speeds were determined by a linear fit. For each $\Rey$, we
averaged the measurement over at least 50 realisations.

\subsection{Numerical simulations}

We consider the motion of incompressible fluid driven through a
circular pipe with a fixed mass-flux. Normalising lengths with the diameter $D$ and 
velocities with the bulk velocity $U_\text{bulk}$, the Navier-Stokes equations read
\begin{equation*}
 \frac{\partial\bm u}{\partial t}+{\bm u}\cdot\bm{\nabla}
{\bm u}=-{\bm{\nabla}p}+\frac{1}{\Rey}\Delta{\bm u}, \;
\hspace{5 mm}\bm{\nabla}\cdot{\bm u}=0
\label{Navier-stokes}
\end{equation*}
where $\bm u$ is the velocity of the fluid and $p$ the pressure. 
The Reynolds number is $\Rey = U_\text{bulk} D/\nu$, where $\nu$ is the kinematic viscosity. 
These equations were solved in cylindrical coordinates $(r, \theta, z)$ 
using a code developed by Ashley~P.~Willis~\cite{Willis2009}, employing a spectral-finite difference method with no-slip boundary conditions 
at the pipe wall $\bm u(1/2,\theta,z,t)=\bm 0$ and periodicity 
in the axial direction. The pressure term was eliminated from the equations 
by using a Toroidal-Poloidal potential formulation of the velocity field,
in which the velocity is represented by toroidal 
$\psi$ and poloidal potentials $\phi$, such that 
$\bm u=\bm{\nabla}\times(\psi\bm{\hat z})+
\bm{\nabla}\times\bm{\nabla}\times(\phi\bm{\hat z})$. 

After projecting the curl and double curl of the Navier-Stokes equations on
the $z$-axis, a set of equations for the potentials $\psi$ and $\phi$ is
obtained. A difficulty, due to the coupled boundary conditions on the
potentials, is solved with an influence-matrix method. In
the radial direction spatial discretization is performed using finite
difference method with a 9-point stencil. Assuming 
periodicity in azimuthal and axial directions, the unknowns, i.e., the
potentials, are expanded in Fourier modes,
\begin{equation*}
A(r,\theta,z,t)=\sum_{k=-K}^{K}\sum_{m=-M}^{M}\hat{A}_{k,m}(r,t)
e^{(i\alpha kz+im\theta)}
\label{equ:Fourier}
\end{equation*}
where $\alpha k$ and $m$ give the wavenumbers of the modes in the
axial and azimuthal directions respectively, $2\pi/\alpha$ fixes the
pipe length $L_z$, and $\hat{A}_{k,m}$ is the complex Fourier
coefficient of mode $(k,m)$. The time-dependent equations are
integrated in time using a second-order predictor-corrector scheme
with a dynamic timestep size, which is controlled using information
from a Crank-Nicolson corrector step. The nonlinear term is evaluated
using a pseudo-spectral technique with the de-aliasing
$\frac{3}{2}$-rule. With expansion~\eqref{equ:Fourier}, the resulting
linear differential equations for the potentials $\psi$ and $\phi$
decouple for each $(k,m)$ mode. This are solved using an LU
decomposition of the resulting banded matrices.  See~\cite{Willis2009}
for more the details of the formulation and solution.

Initial conditions were prepared at $\Rey=2000$ in 133$D$ and 180$D$
pipes for simulations at $\Rey>2000$. At low $\Rey
\gtrsim 2000$ puff-splitting is extremely unlikely~\cite{kavila2011} and puffs remain approximately constant in length (about $20D$) as they travel downstream along the
pipe. Hence simulations at $\Rey=1910,1920,2000$ were carried out in a
shorter $24\pi\approx75D$ pipe, with initial conditions prepared at
$\Rey=1950$. In Table~\ref{tab:Re_reso} the length of the pipes and
numerical  resolutions used at each Reynolds number are listed.

\begin{table}[h!]
\centering
\begin{tabular}{p{1cm}p{1cm}p{1cm}p{1cm}p{1cm}|p{1cm}p{1cm}p{1cm}p{1cm}p{1cm}}
\hline
$\Rey$ & $L_z (D)$ & $N$ & $K$ & $M$ & $\Rey$ & $L_z$ & $N$ & $K$ & $M$\\ 
\hline
1910 & $24\pi$ & $48$ & $640$  & $32$ & 3000 & $180$   & $72$ & $2560$ & $48$\\
1920 & $24\pi$ & $48$ & $640$  & $32$ & 3200 & $180$   & $72$ & $2560$ & $54$\\
2000 & $24\pi$ & $48$ & $768$  & $40$ & 3500 & $180$   & $72$ & $2560$ & $54$\\
2200 & $133$ & $48$ & $768$  & $40$ & 3750 & $133$   & $72$ & $2048$ & $54$\\
2300 & $133$   & $64$ & $1536$ & $40$ & 4000 & $133$   & $72$ & $2048$ & $54$\\
2400 & $133$   & $64$ & $2048$ & $48$ & 4500 & $180$   & $80$ & $3072$ & $64$\\
2800 & $180$   & $72$ & $2560$ & $48$ & 5000 & $180$   & $80$ & $3072$ & $64$\\
2600 & $133$   & $64$ & $2048$ & $48$ & 5500 & $180$   & $96$ & $3840$ & $80$\\
\hline
\end{tabular}
\caption{\label{tab:Re_reso} The domain size and resolution for the simulation
  at all the Reynolds numbers we considered. Note that in physical space there
  are $3K$ and $3M$ grid points in axial and azimuthal directions.}
\end{table}

The fronts were detected by setting an appropriate cut-off. In this paper, the
local intensity was computed as
\[ \int\int{(u_r^2+u_{\theta}^2)}rdrd\theta,\]
and a cut-off of $5\times10^{-4}$ was chosen for all the simulations
to determine the position of laminar-turbulent fronts. We tested
different cut-off values and found that the front speed was
insensitive to them.

The expansion speed of the downstream front was found to accelerate
substantially during the initial stages of the simulation.  In order to obtain the asymptotic value of the speed, we determined the length of the turbulent region $L_0$ beyond which the speed statistics become length-independent. We found that for $\Rey<4000$,  $L_0>60D$  was sufficient, whereas for $\Rey\ge 4000$  $L_0>100D$ was required. This is the reason why very long pipes as reported in Table~\ref{tab:Re_reso} were used. At each $\Rey$ the speed was determined by computing $(z_\text{end}-z_0)/(t_\text{end}-t_0)$ for each run and then averaging over a total of 20 runs. The initial time $t_0$ corresponds here to the time at which the turbulent region has reached the length $L_0$.

\section{Theory}

\subsection{Model details}

The model is a two-component system of advection-reaction-diffusion equations
\begin{equation}
\frac{\partial q}{\partial t} + (u-\zeta) \frac{\partial q}{\partial x} 
= f(q,u) +  D\frac{\partial^2 q}{\partial x^2}, \quad
\frac{\partial u}{\partial t} + u \frac{\partial u}{\partial x} 
= \epsilon g(q,u) 
\label{eq:full} 
\stepcounter{equation}\tag{S\theequation}
\end{equation}
where $q$ represents the level of turbulent fluctuations and $u$ the axial
velocity on the centreline. 
The nonlinear reaction functions $f$ and $g$ are given by
$$
f(q,u) = q \left( r + u - 2 - (r + 0.1) (q - 1)^2  \right), \quad
g(q,u) = 2 - u + 2 q (1 - u)
$$
where the parameter $r$ corresponds to a suitably scaled Reynolds number. 

The model and the role of the fitting parameters ($D$, $\zeta$ and $\epsilon$)
are most easily understood by first considering the equations in the absence
of spatial derivatives. In this case the model reduces to the ordinary
differential equations (ODEs)
$$
\dot q = f(q,u), \qquad  \dot u = \epsilon g(q,u) 
$$
These ODEs are the core of the model as they describe the interaction between
the turbulent fluctuations $q$ and axial velocity $u$ locally in space.  The
functional forms are designed to qualitatively capture the well-established
physics of this interaction~\cite{Coles:1962,wygnanski1973} with minimal
nonlinearities.  (In a previous approach~\cite{barkley2011}, the variable $u$
corresponded to the axial velocity of pipe flow in the frame of reference
moving at the bulk velocity $U_\text{bulk}$, here $u$ corresponds to velocity
in the lab frame so that $u=2$ for laminar flow.)

The nullclines for the ODEs are given by $f(q,u)=0$ and $g(q,u)=0$.  For all
parameter values of the ODEs these nullclines intersect at the fixed point $(u=2,
q=0)$ corresponding to laminar, Hagen--Poiseuille flow. By design $\epsilon$
sets ratio of the time scale of $u$ relative to $q$.  (Previously~\cite{barkley2011},
parameters $\epsilon_1$ and $\epsilon_2$ appeared in the model. Here we have a
simplified the model to a single time-scale ratio $\epsilon$. In terms of
$\epsilon$, the previous parameters would be $\epsilon_1 = \epsilon$ and
$\epsilon_2 = 2 \epsilon$.)

Now consider the full model equations. In addition to the local terms given by
$f$ and $g$, the model has first and second spatial derivatives.  The first
derivative terms account for nonlinear advection in the streamwise
direction. For the $u$ equation we use the advective nonlinearity following
directly from the Navier--Stokes equations. For the turbulent field $q$, the
parameter $\zeta$ accounts for a diminished advection of $q$ in comparison
with the centreline velocity $u$. The streamwise velocity is maximal on the
centreline and the turbulent field is not advected at this speed.  We have
simulated turbulent flow in short pipes ($L= 12D$) and have verified that
turbulent structures are advected considerably more slowly than the centreline
velocity. This effect involves complex processes in the pipe cross-section. We
include in the model the simplest term that can account for diminished
advection. (Previously~~\cite{barkley2011}, the model contained only linear
advection, the fixed difference in the advection of the $q$ and $u$ fields was
expressed by an additional first-derivative term on the right-hand-side of the
$u$-equation. This effectively corresponded to $\zeta=1$ in the current
model.)  We return to the importance of the parameter $\zeta$ after we
  derive expressions for front speeds in the model.

The diffusive term in equation \eqref{eq:full} accounts for the processes by
which a region of turbulence flow couples to and thereby excites adjacent
laminar flow. The physical processes involved are complex and not fully
understood
\cite{wygnanski1975,wygnanski1973,Duguet:2010,moxey2010,kavila2011,holzner2013}.
The second-derivative is the most natural choice for modelling such a
coupling. The coupling strength, or diffusion coefficient, $D$ is the final
model parameter.

\subsection{Asymptotic analysis}

The asymptotic analysis follows very closely that of~\cite{Tyson:1988}. Let
the three roots of $f$ be noted $q^0, q^\pm$. The laminar branch is $q^0 = 0$
for all $u$ and $r$, while the upper and lower branches $q^\pm$ are functions
of $u$ and $r$. The laminar $q^0$ and upper $q^+$ branches are stable. For
small $\epsilon$ the dynamics of the system separates into slow regions and
fast front regions. In the slow regions the system is slaved to one of the
stable branches (slow manifolds) and $u$ evolves on a slow scale, e.g.\ along
the upper branch $q^+$
$$
q = q^+(u), \qquad 
\frac{\partial u}{\partial t'} + u \frac{\partial u}{\partial x'} 
= g(q^+(u),u) 
$$
where $x'= \epsilon x$ and $t'= \epsilon t$ are slow scales.  

In the fast regions, fronts are formed as the system transitions between the
stable branches, from $q^0$ to $q^+$ as $x$ increases for a upstream front
while $q^+$ to $q^0$ for a downstream front. Let $c$ denote the speed of the
front and go into a frame of reference moving at speed $c$. Locate the now
stationary front at $x=0$ and work in an inner (stretched) variable.  At
leading order in $\epsilon$ the equations in the stretched coordinate become
\begin{equation}
q'' + s q' + f(q,u) = 0, \label{eq:qfast} 
\stepcounter{equation}\tag{S\theequation} \quad
\end{equation}
\begin{equation}
u' = 0 \label{eq:ufast} 
\stepcounter{equation}\tag{S\theequation} 
\end{equation}
where 
$$
s \equiv \frac{c - (u_f - \zeta) }{\sqrt{D}}  
$$
From Eq.~\eqref{eq:ufast}, at leading order $u$ is constant across a
front. Call this constant value $u_f$.
Equation \eqref{eq:qfast} must be solved subject to boundary conditions.  For
a downstream front these are
\begin{equation}
q(-\infty) = q^+(u_f), \quad q(+\infty) = q^0 
\label{eq:BC_down}
\stepcounter{equation}\tag{S\theequation} 
\end{equation}
For an upstream front the boundary conditions are reversed, but this can be
accounted for by a change of sign of $s$ in Eq.~\eqref{eq:qfast}. 

\begin{figure}[h]
\begin{center}
\includegraphics[width=4.2in]{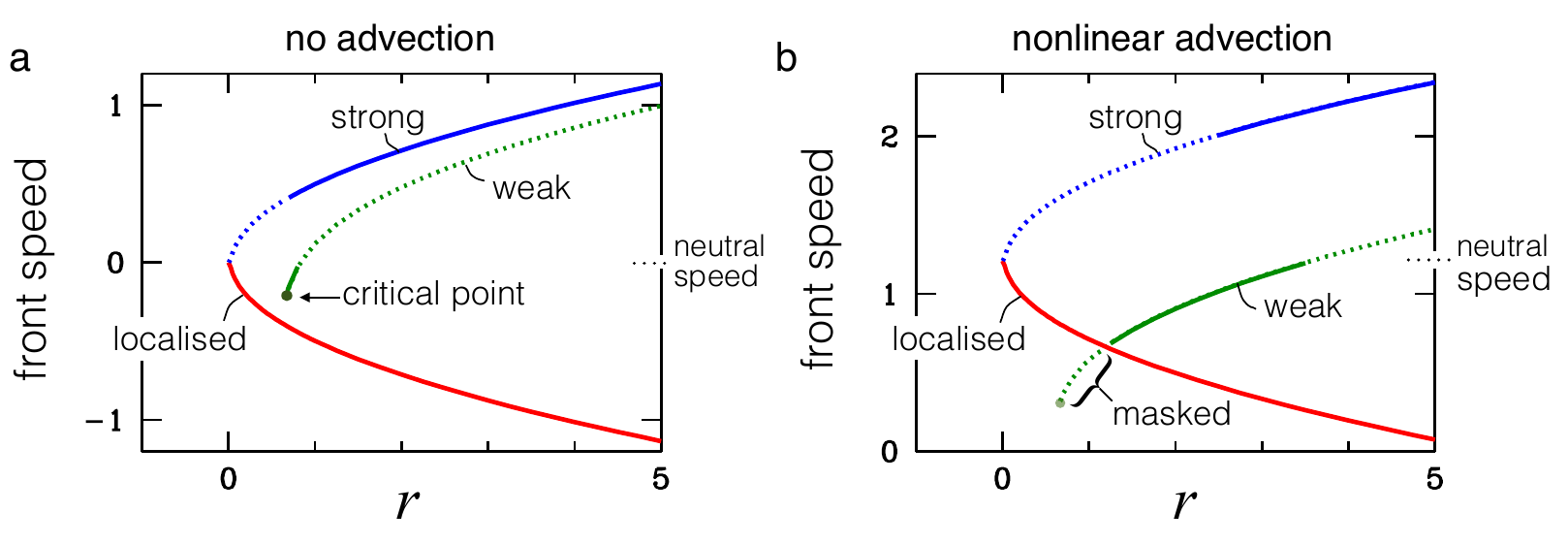} 
\end{center}
\caption{{\bf Speed of model fronts in the asymptotic limit of sharp fronts.}
  Speeds as a function of model Reynolds number $r$ both {\bf a} without and
  {\bf b} with advection.
   While strong downstream fronts cannot exist and have no physical meaning
   below the formation of the upper branch fixed point, the expression for
   strong front speeds \eqref{eq:cstrong} still gives the speed such a strong
   downstream front would have and these speeds are shown as dashed.
  The effect of nonlinear advection in {\bf b} is to mask the
  nominal critical point for the onset of fully turbulent flow. 
  The neutral speed is naturally displaced from the bulk speed 
  $U_\text{bulk} = 1$. 
}
\label{fig:fronts_SI} 
\end{figure}

In summary, the speed of a front at given value of $u = u_f$ is found by
solving
\begin{equation}
q'' + s q' + f(q,u_f) = 0 \label{eq:BVP}
\stepcounter{equation}\tag{S\theequation} 
\end{equation}
subject to boundary conditions \eqref{eq:BC_down}. This will give a value of
$s$, which will depend on $u_f$ and $r$. Denote it $s(u_f, r)$. From this the
front speed is
\begin{equation}
c =  u_f - \zeta  \pm \sqrt{D} \, s(u_f, r) \label{eq:cgen}
\stepcounter{equation}\tag{S\theequation} 
\end{equation}
with $+$ for a downstream front and $-$ for an upstream front. 
For the strong downstream front and all upstream fronts, $u_f = 2$. Hence
their speeds are
\begin{equation}
c =  2 - \zeta  \pm \sqrt{D} \, s(2,r)  \label{eq:cstrong}
\stepcounter{equation}\tag{S\theequation} 
\end{equation}
For the weak downstream front $u_f = u_{ss}$ where $u_{ss}$ is the upper
branch steady state. Hence
\begin{equation}
c =  u_{ss} - \zeta  + \sqrt{D} \, s(u_{ss},r)   \label{eq:cweak}
\stepcounter{equation}\tag{S\theequation} 
\end{equation}
Figure \ref{fig:fronts_SI} shows model front speeds as a function of model
Reynolds number.  Figure \ref{fig:fronts_SI}b is the same as Fig.~2g in the
main paper, except over a smaller range of $r$.
Speeds are from Eqs.~\eqref{eq:cstrong} and \eqref{eq:cweak}. 

Note that the neutral speed in the model is $2 - \zeta$. This follows
immediately from Eq.~\eqref{eq:cstrong} where one can see that the upstream
speed (minus sign) and strong downstream speed (plus sign) are symmetric with
respect to $2 - \zeta$. This is the advection speed of turbulence in the
absence of front dynamics due to transitions between laminar and turbulent
flow. Without the parameter $\zeta$, the neutral speed would be the maximum
centreline velocity.  This is neither consistent with the observed neutral
speed, nor is it reasonable that turbulent structures would be advected at the
{\em maximum speed found in the flow}.

Figure~\ref{fig:fronts_SI}a shows fronts speeds without the inclusion of
advection terms (first derivatives in $x$) in the model equations.
Without these terms the front speeds become
$$
c =  \pm \sqrt{D} \, s(2,r)  
$$
for the strong downstream front and all upstream fronts, and 
$$
c =  \sqrt{D} \, s(u_{ss},r)
$$
for the weak downstream front.
The transition to expanding turbulence is discontinuous.
Including linear advection (as was done previously~\cite{barkley2011}) will
result in an overall shift in all front speeds, and can affect the asymptotic
stability of branches (technical details on stability for the asymptotic
branches will be presented elsewhere) but will not change the discontinuous
nature of the transition.

This highlights the role of nonlinear advection in the bifurcation scenario:
without the physical affect of nonlinear advection the weak front branch has a
distinct critical point and the transition to expanding turbulence is
first-order (discontinuous). 

Note that in Figs.~2a, 2b, and 2c, of the main paper, solutions $q(x)$ are
obtained directly from the full model equations at $\epsilon = 0.01$, which is
sufficiently small that these solutions are visually close approximations to
the $\epsilon \to 0$ limit. Figures~2d, 2e, 2f, show the nullclines for
exactly the cases shown in Figs.~2a, 2b, and 2c; however, the phase portraits
are sketches. This both facilitates the colouring of the front branches and
also, even at $\epsilon = 0.01$, the jumps between the branches of $q$ are not
seen as completely vertical in the phase plane.

It is well established that exact coherent structures in pipe flow lie along
upper and lower branches. Most known exact solutions are spatially extended,
in the form of travelling waves~\cite{kerswell2005,eckhardt2007, gibson2008,
  kawahara2011}, but recently spatially localised states were also
found~\cite{avila2013b,chantry2014}. The model captures these states in a
minimal way. The fixed points (one stable and one unstable) arising as the
model transitions to bistability can be viewed as upper and lower branches of
spatially extended travelling-wave solutions. The cubic nonlinearity in $f$ is
that minimally required for this separation into upper and lower branch
states. The model also has localised states (puffs) and importantly unstable
small-amplitude localised solutions (not discussed in this paper, but see
Refs.~\cite{barkley2011,Rinzel:1982,Tyson:1988}) corresponding to edge states,
both in the puff regime and in the fully turbulent regime.

\begin{figure}[h]
\begin{center}
\includegraphics[width=4.5in]{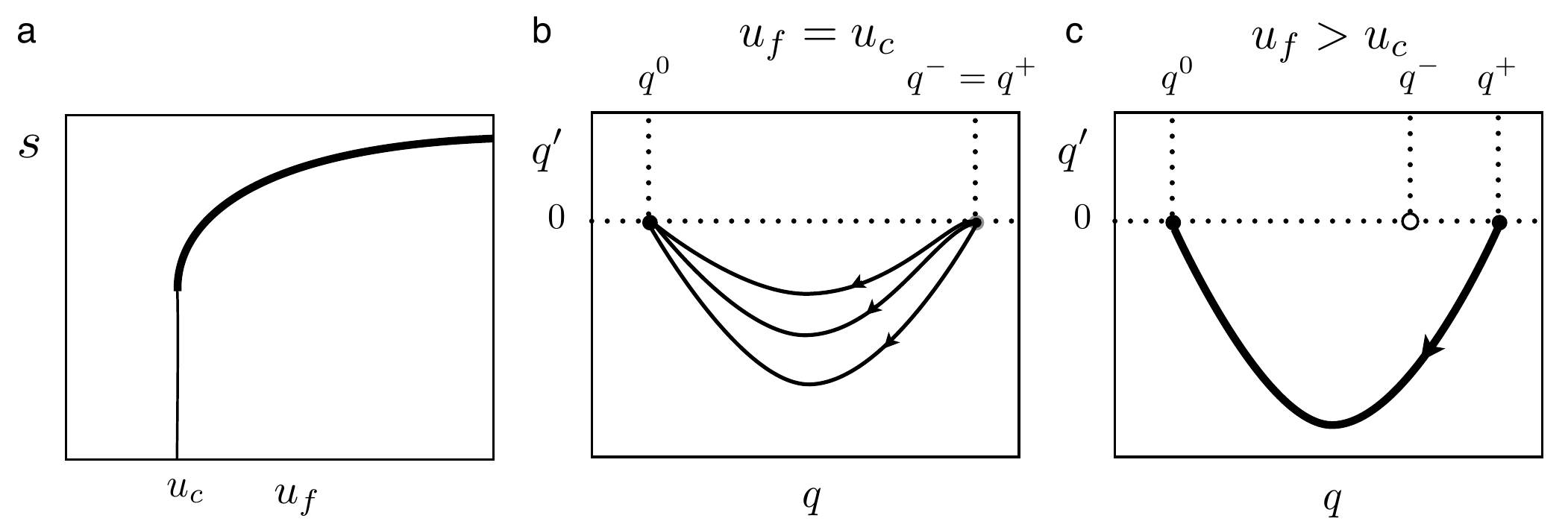} 
\end{center}
\caption{{\bf Front speeds at critical point.}  Sketch illustrating solutions
  to the boundary value problem \eqref{eq:BVP} for a downstream front near the
  critical point. {\bf a}, eigenvalue $s$ as a function of $u_f$. $u_c$ is the
  value of $u_f$ such that $q^-=q^+$. For this value there are infinitely many
  possible eigenvalues $s$ indicated by a thin line.  {\bf b,c} phase planes
  $(q, q')$ showing solutions for the second order differential equation
  \eqref{eq:BVP}. Downstream fronts are heteroclinic connections from the
  upper fixed point $q^+$ to the lower fixed point $q^0$. When $u_f=u_c$ and
  hence $q^-=q^+$, the upper fix point is not hyperbolic and there are
  infinitely many connections, each corresponding to a value of $s$. When
  $u>u_c$, $q^+$ is hyperbolic and there is a unique connection and hence a
  unique solution $s$. }
\label{fig:critical_point_SI} 
\end{figure}

Finally, we comment on what takes place at the critical point. As with all
materical in this section, the discussion 
follows closely Refs.~\cite{Rinzel:1982,Tyson:1988}.
Figure~\ref{fig:critical_point_SI} illustrates solutions to the boundary value
problem \eqref{eq:BVP} in the case of a downstream front. For a fixed value of
$r$, the eigenvalue $s$ and solution $q$ depend on $u_f$, the value of $u$ at
the front. Note in particular that the boundary conditions \eqref{eq:BC_down}
depend on $u_f$.

Downstream fronts are heteroclinic connections from $q^+$ to $q^0$, where
``time'' in the phase plane corresponds to space $x$.
The phase plane is two-dimensional, coordinates $q$ and $q'$, because 
Eq.~\eqref{eq:BVP} is a the second order differential equation. 
As illustrated in Fig.~\ref{fig:critical_point_SI}c, for generic $u_f$, both
$q^+$ and $q^0$ are hyperbolic fixed point (saddles) in the phase plane and
hence a heteroclinic connection exists only for a unique value of $s$. This
determines $s$ as a function of $u_f$, as shown by the bold curve in
Fig.~\ref{fig:critical_point_SI}a. However, when $u_f$ takes the value such
that $q^+ = q^-$, i.e.\ at the nose of the $q$ nullcline, then the upper fixed
point is no longer hyperbolic and there exist infinitely many heteroclinic
connections from $q^+$ to $q^0$, and hence infinitely many possible values of
$s$. These appear as the thin line in Fig.~\ref{fig:critical_point_SI}a. 

Now, as the parameter $r$ is varied (as in Fig.~2 of the main paper), the
nullclines vary. The critical point is where upper branch steady state occurs
at the limit point of the $q$ nullcline, that is the upper fixed point is at
$q^+=q^-$. For $r$ smaller than this value, the downstream front can take any
of an infinite range of values because the downstream front can occur at a
value of $u_f = u_c$. 
For a puff, this infinite range of possible values is the mechanism that
allows the speed of the downstream front to select the same value as for the
upstream front. As a result puffs remain localised while traveling along the
pipe.
However, for $r$ larger than the critical value the
upper branch fixed point no longer permits downstream fronts to occur at $u_f
= u_c$, as seen in (as in Fig.~2e of the main paper). This restricts the
possible values of $s$, and hence possible speeds of the downstream front, to the
bold portion of the branch illustrated in Fig.~\ref{fig:critical_point_SI}a.
Hence, as $r$ passes through the critical point there is an abrupt change in
the allowed values of the downstream front speeds, from an infinite to a
finite range. Without nonlinear advection, the abrupt change is manifested as
a discontinuous change in the speed of the downstream front. With nonlinear
advection, there is still an discontinuous change to the allowed values of
$s$, but the speeds are smaller than those of the upstream front, so the
discontinuity in allowed solutions is masked.

\subsection{Collapsing pipe and duct data}

\begin{figure}
\begin{center}
\includegraphics[width=5.0in]{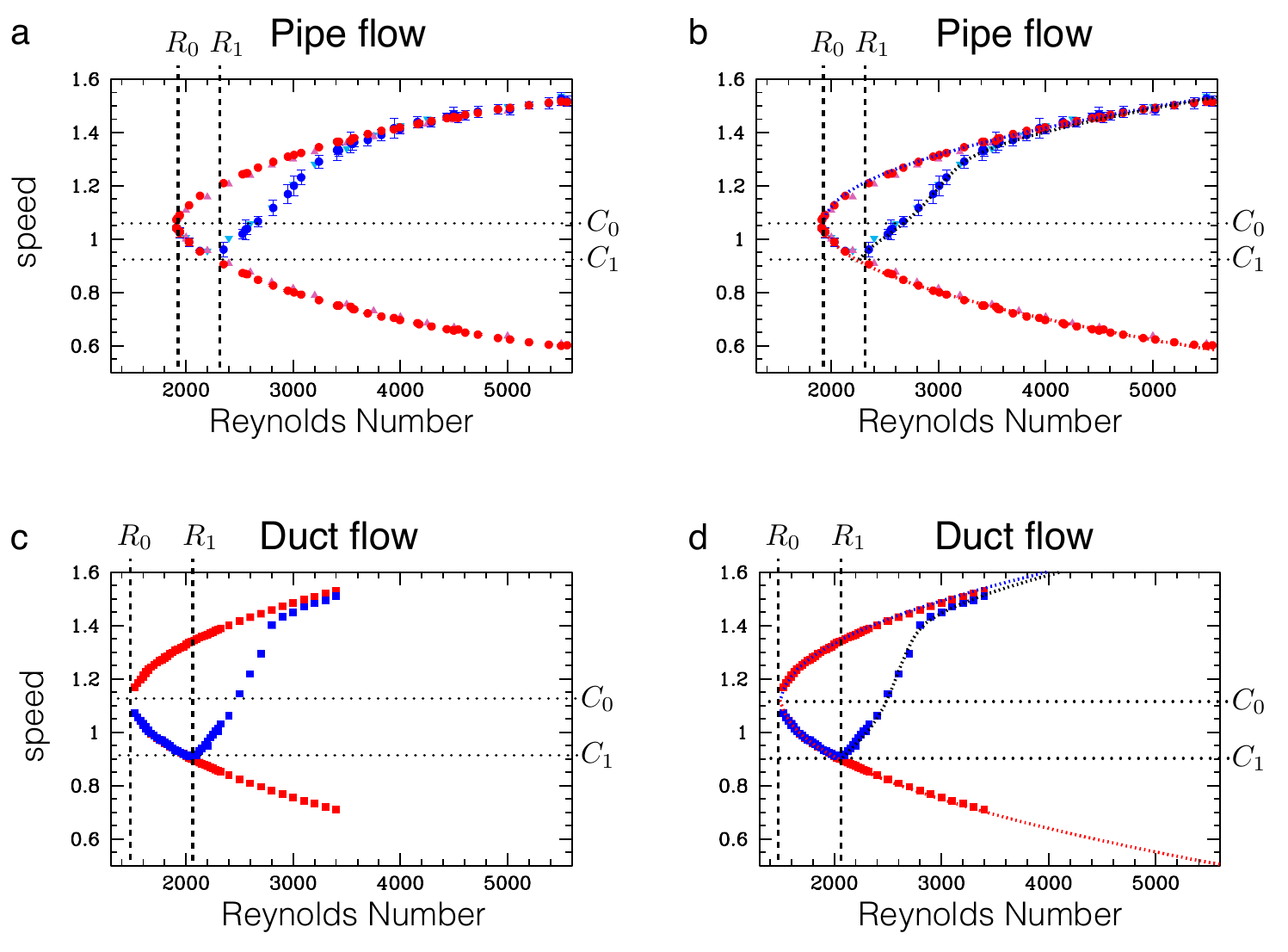} 
\end{center}
\caption{{\bf Determination of corresponding Reynolds numbers and speeds for
    pipe and duct flow.} Speeds from pipe and duct flow are plotted, as in
  Fig.~1c of the main paper, but additionally with the upstream front speeds
  reflected about the neutral speed $C_0$. Left column shows only experimental
  and simulation data while the right column includes also model fits to the
  data.
The determined values for $R_0$, $R_1$, $C_0$, $C_1$, are:
$R_0 = 1920$, $C_0 = 1.06$ $R_1 = 2250$, and $C_1 = 0.92$ for pipe flow, and
$R_0 = 1490$, $C_0 = 1.12$ $R_1 = 2030$, and $C_1 = 0.90$ for duct flow.  
}
\label{fig:reduced_Re_SI} 
\end{figure}

In order to collapse data from pipe and duct flow onto a single plot, it is
necessary to determine specific Reynolds numbers and speeds from measured data
(see figure \ref{fig:reduced_Re_SI}) which will then be used to align the data
from the two flows.
We stress that while the procedure is informed from the model analysis, it
requires only measured data and the same procedure could be applied to data
from other shear flows.

Figures \ref{fig:reduced_Re_SI} {\bf a} and {\bf c} show data from pipe and
duct flow, respectively, plotted with the upstream speeds additionally
reflected about the neutral speed, here labelled $C_0$. The value of $C_0$ is
determined to be that for which reflected upstream data coincides with the
downstream data at sufficiently large Reynolds number. Figures
\ref{fig:reduced_Re_SI} {\bf b} and {\bf d} show the same data, but with model
speeds (determined subsequently) also plotted to additionally guide the
eye. In the case of pipe flow is it possible to determine $C_0$ to better that
2\% accuracy by the procedure. For duct flow, our estimation is that the
downstream front speed has not quite reached the reflected upstream speed at
the highest Reynolds number accessible to present experiments. Nevertheless,
$C_0$ is still quite well determined by this procedure.
From the same plots, the value of the Reynolds number $R_0$ at which the
upstream front obtains the neutral speed $C_0$, is easily determined.  

Then, from the data, we determine the Reynolds number $R_1$ where the
downstream weak front first deviates from the downstream front. Again, this
can in principle be determined solely from the data, although using model
fits to the weak branch can give further confidence in the determined values. 
$C_1$ is the value of the front speed at $R_1$.  

\begin{figure}[h]
\begin{center}
\includegraphics[width=3.0in]{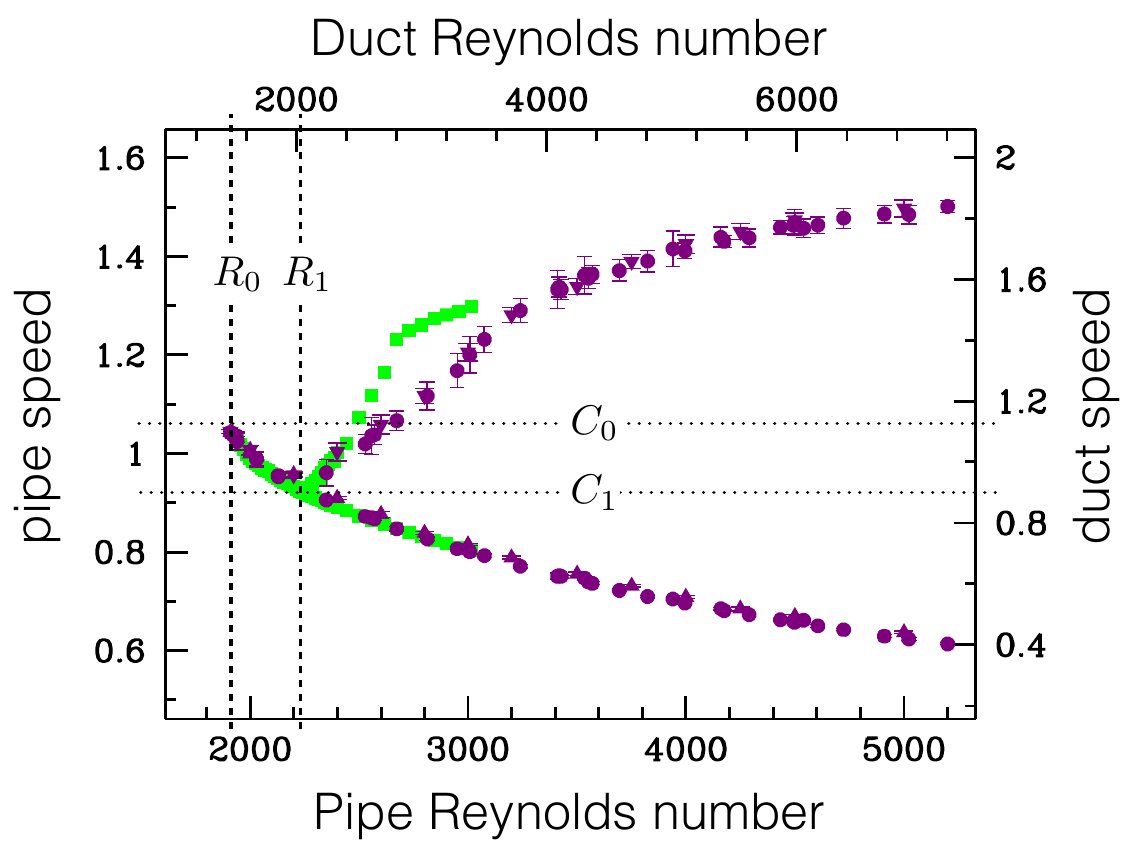} 
\end{center}
\caption{{\bf Collapse of pipe and duct data.} Pipe and duct flow are plotted
  together using different axes.  The data are plotted such as to align the
  two points $(R_0, C_0)$ and $(R_1, C_1)$ for each data set, e.g. $(R_1, C_1)
  = (2250, 0.92)$ for pipe flow is aligned with $(R_1,C_1) = (2030, 0.90)$ for
  duct flow, bringing into alignment the onset of weak front. }
\label{fig:collapse_SI} 
\end{figure}

Once the values $(R_0, C_0)$ and $(R_1, C_1)$ have been found for each flow,
the data can be collapsed by plotting each data set to align the two points
$(R_0, C_0)$ and $(R_1, C_1)$, as seen in figure \ref{fig:collapse_SI}. This
is equivalent to simply choosing the origin and scaling for the axes for the
two flows. The upstream and strong downstream fronts each collapse, while the
weak-front branch does not.
        
\subsection{Determining model parameters}

There are three model parameters, namely $D$, $\zeta$, and $\epsilon$, to be
determined to quantitatively relate the model speeds to the measured data for
each flow.

The generic model cannot be expected to predict the flow specific values $R_0,
R_1, C_0$, and $C_1$, and moreover there is nothing universal about these
values.  Instead, given these flow-specific values, the model is expected to
capture the form of the various branches seen in the collapsed data of
Fig.~\ref{fig:collapse_SI}.
For fitting model parameters it is useful to plot the collapsed data in terms
of reduced Reynolds number and reduced speed
\begin{equation}
\frac{R - R_0}{R_1 - R_0}, \qquad 
\frac{1}{2} \frac{C - C_0}{C_0 - C_1}
\label{eq:reduced_def}
\stepcounter{equation}\tag{S\theequation} 
\end{equation}
which amounts only to a relabelling of the axes in Fig.~\ref{fig:collapse_SI}
to shift the neutral speed to zero and scale the onset of the weak front to
the point $(1,-1/2)$.
%
%
%
As will apparent momentarily, the reason for including 1/2 in the reduced speed
is that model speeds are typically about half those of the reduced speeds for
experimental data.
 
\begin{figure}
\begin{center}
\includegraphics[width=5.5in]{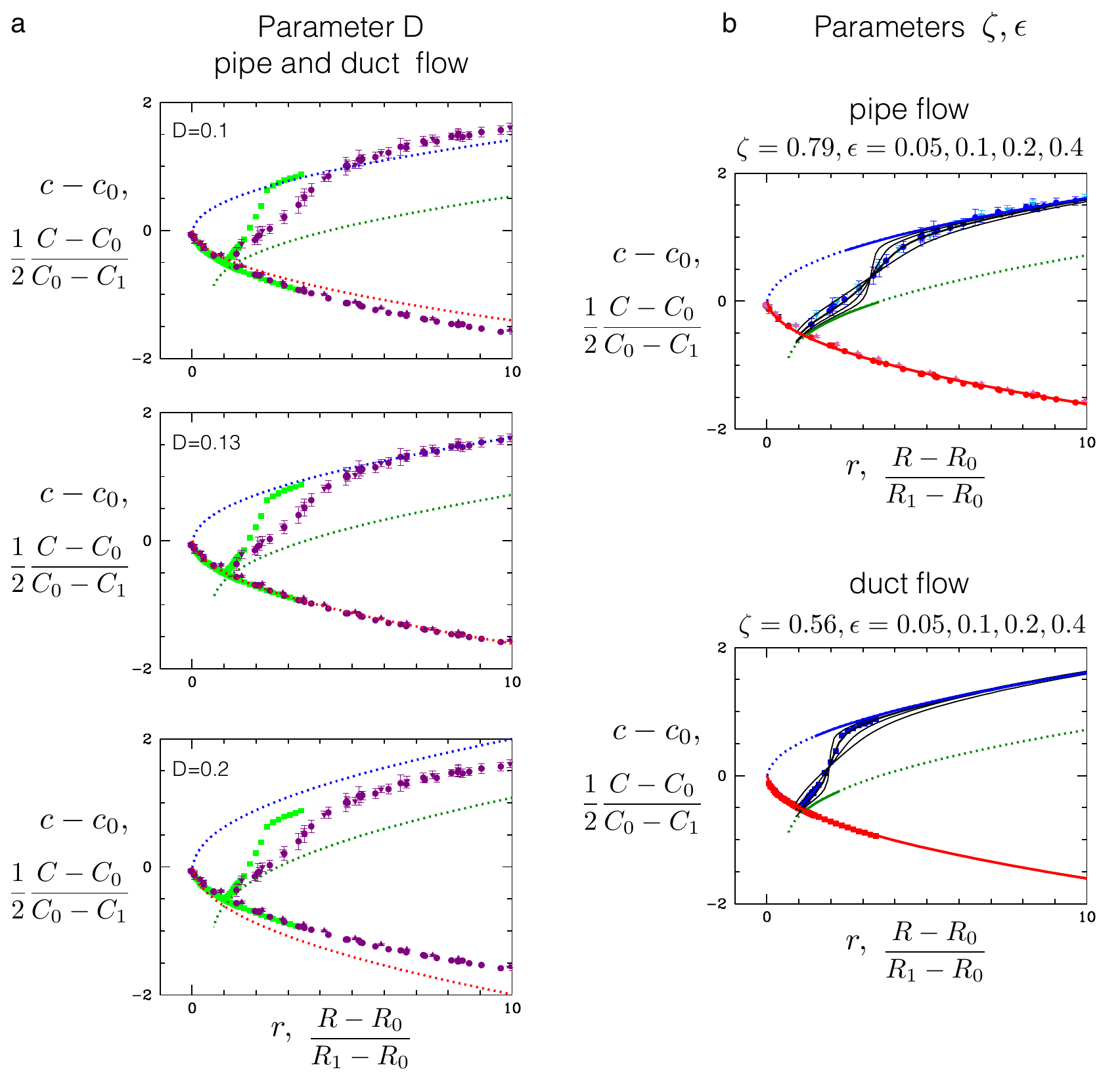} 
\end{center}
\caption{{\bf Determination of model parameters for pipe and duct flow.}  {\bf
    a}, Determination of $D$. Points are collapsed data from pipe and duct
  flow (as in Fig.~\ref{fig:collapse_SI}) here plotted in terms of reduced
  Reynolds number and reduced speed. Dotted curves are asymptotic speed curves
  (as in Fig.~\ref{fig:fronts_SI}) plotted in term of model Reynolds number
  $r$ and speed $c - c_0$. For $D=0.13$ there is very good agreement between
  data and model. This choice fixes the asymptotic branches (dashed lines).
{\bf b}, Determination of $\zeta$ and $\epsilon$. Pipe and duct flow are
necessarily considered separately. In each case, downstream branches are shown for four values of
$\epsilon$. Smaller values give more abrupt transition between weak and strong
branches. 
}
\label{fig:fitting_SI} 
\end{figure}

We first consider the value of the parameter $D$. 
%
%
%
%
We select $D$ so as to fix a simple relationship between model and measured
quantities for both flows.  Specifically, in Fig.~\ref{fig:fitting_SI}a we
plot the collapsed pipe and duct data together with the asymptotic results
from the model for different values of $D$.  The model results are plotted
directly in terms of model Reynolds number $r$ and model speed $c-c_0$ (the
model speed $c$ shifted by the model neutral speed $c_0$). For $D=0.13$ the
upstream front and strong branches match the collapsed experimental data
extremely well.  Note, the strong and weak asymptotic curves in
Fig.~\ref{fig:fitting_SI}a are independent of the other two model parameters,
$\epsilon$ and $\zeta$.

The result is that using only one parameter, $D$, and fixing its value to
$0.13$, the model does not only fit very well the upstream and strong downstream
front speeds for both flows, but also a simple relationship between
model and experimental data is fixed, namely
\begin{equation}
r = \frac{R - R_0}{R_1 - R_0}, \qquad 
c - c_0 = \frac{1}{2} \frac{C - C_0}{C_0 - C_1}
\label{eq:reduce_relation}
\stepcounter{equation}\tag{S\theequation} 
\end{equation}
Given the flow specific values $R_0$, $R_1$, $C_0$, and $C_1$,
Eqs.~\eqref{eq:reduce_relation} can be inverted to obtain Reynolds number $R$
and speed $C$ from the model values $r$ and $c$. This is how model results are
mapped to Reynolds number and speed in Fig.~3 of the main paper.

The remaining two model parameters dictate the behaviour of the downstream
fronts as they transition from weakly expanding to strongly expanding. Here
pipe and duct flow differ and so necessarily the values of the fitting
parameters will be different for the two flows. 
See Fig.~\ref{fig:fitting_SI}b. 

The value of $\epsilon$ dictates how quickly the system jumps from the weak to
the strong branch. Large values give smoother transitions while smaller values
give more abrupt transitions. 
The value of $\zeta$ dictates how long the system follows the weak branch
before transitioning to the strong branch.  Larger values, as for pipe flow,
result in a delay in transition, while smaller values of $\zeta$, as for the
fit to duct flow, result in more immediate transition.
We did not apply a formal procedure for determining $\zeta$ and $\epsilon$ for
each of the flows. Rather they were determined simply by eye. In both cases it
is quite easy to adjust $\zeta$ and $\epsilon$ so that the transition from
weak to strong front follows the measured data.

\section{Control}

\begin{figure}
\begin{center}
\includegraphics[width=4.0in]{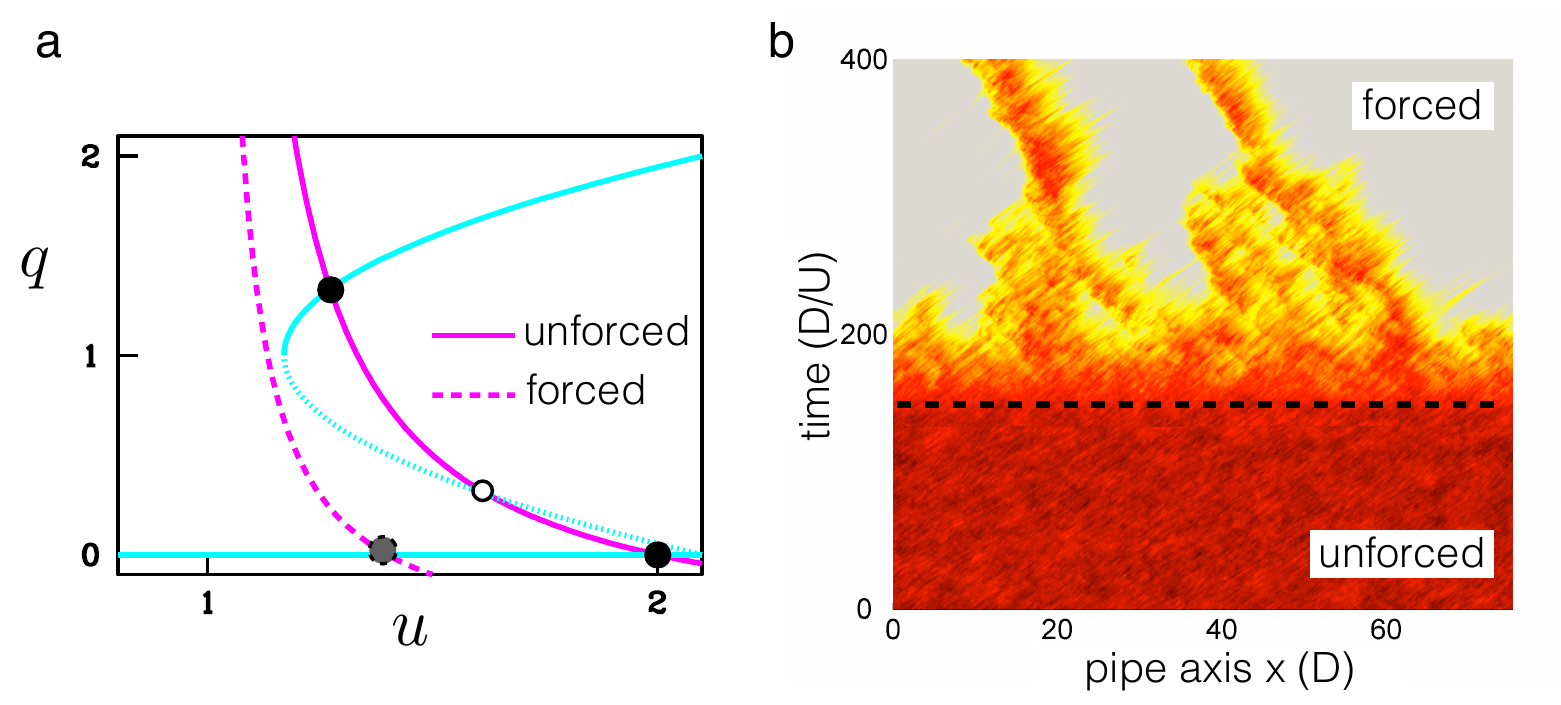} 
\end{center}
\caption{{\bf Illustration of control based on removing turbulent fixed
    point.}  {\bf a}, control concept illustrated in the model phase
  plane. Without forcing (i.e.\ without control), there is an upper branch
  fixed point (upper intersection of nullclines) corresponding to fully
  turbulent flow. Applying an additive forcing term to the $u$-equation
  corresponds to forcing the shear profile and blunting its shape. This can
  remove the turbulent fixed point thus eliminating fully turbulent flow. 
{\bf b}, Proof of concept in a direct numerical simulation of pipe flow
at $R = 5000$. Without forcing the flow is fully turbulent.  A global body
force is applied and blunts the velocity profile to a more plug-like form.
Subsequently, only localised turbulent patches remain,
reminiscent of those at much lower $R$.  }
\label{fig:control_SI} 
\end{figure}

The model suggests that the fully turbulent state can be destabilized by
removing the upper turbulent fixed point as depicted in
Fig.~\ref{fig:control_SI}a. In the model this is achieved by forcing the
variable $u$ corresponding to the state of shear profile. The reduction of $u$
by forcing corresponds to a blunting of the shear profile.

To demonstrate that the fully turbulent state can indeed be destabilized by
removing the turbulent fixed point, as suggested by the model, we have
performed a direct numerical simulation of pipe flow at $R=5000$. Initially the forcing is not applied and the flow is fully turbulent.  Starting at time 175 a
global body force is gradually switched on (fully applied by time 200) which
blunts the velocity profile to a more plug-like form (the same forcing is used
as in Hof {\em et al.} \cite{hof2010}).  As can be seen, turbulent intensity
subsequently decreases and eventually the fully turbulent flow destabilizes
and degenerates into localised turbulent patches, just as the natural ones (puffs) at
lower Reynolds number (below $\sim 2300$) in the absence of any additional
force. Further details will be presented elsewhere.

\end{document}